\documentstyle[pjw_cite3,psfig,graphicx]{mn}
\title[SS Cygni in outburst]
{The X-ray and extreme-ultraviolet flux evolution of SS~Cygni
throughout outburst}

\author[P.\,J.\  Wheatley, C.\,W.\  Mauche \& J.\,A.\  Mattei]
{Peter J.\  Wheatley$^1$, Christopher W.\  Mauche$^2$ and Janet A.\  Mattei$^3$ \\
$^1$Department of Physics and Astronomy, University 
of Leicester, University Road, Leicester, LE1 7RH \\
$^2$Lawrence Livermore National Laboratory,
L-473, 7000 East Avenue, Livermore, CA 94550, USA \\
$^3$American Association of Variable Star Observers, 25
Birch Street, Cambridge  MA 02138-1205, USA
}
\def\euve{\it EUVE\rm }
\def\rxte{\it RXTE\rm }

\def\TD-1{\it TD-1\rm }
\def\tkev{\thinspace{ke\kern-.15em V}}
\def\tev{\thinspace{e\kern-.15em V}}

\newcommand\lax{{\lower0.75ex\hbox{ $<$ }\atop\raise0.5ex\hbox{ $\sim$}}}
\newcommand\gax{{\lower0.75ex\hbox{ $>$ }\atop\raise0.5ex\hbox{ $\sim$}}}
\newcommand\ion[2]{#1$\;${\small\rmfamily\@Roman{#2}}\relax}
\begin{document}
\maketitle

\begin{abstract}
We present the most complete multiwavelength coverage of any dwarf nova
outburst: simultaneous optical, {\it Extreme Ultraviolet Explorer\/}, and 
{\it Rossi X-ray Timing Explorer\/} observations of SS Cygni
throughout a narrow asymmetric outburst. Our data show that the high-energy
outburst begins in the X-ray waveband 0.9--1.4\,d after the beginning of the
optical rise and 0.6\,d {\it before\/} the extreme-ultraviolet rise. The 
X-ray flux drops suddenly, immediately before the extreme-ultraviolet
flux rise, supporting the view that both components arise in the boundary
layer between the accretion disc and white dwarf surface. 
The early rise of the X-ray flux
shows the propagation time of the outburst heating wave may have been 
previously overestimated. 

The transitions between X-ray and extreme-ultraviolet dominated emission
are accompanied by intense variability in the X-ray flux, with timescales of 
minutes. As detailed by
Mauche \& Robinson, dwarf nova oscillations are detected throughout the
extreme-ultraviolet outburst, but we find they are absent from the 
X-ray lightcurve. 

X-ray and extreme-ultraviolet luminosities imply accretion rates 
of $ 3\times 10^{15}\,\rm g\,s^{-1}$ in quiescence, $ 1\times 
10^{16}\,\rm g\,s^{-1}$ when the boundary layer becomes optically
thick, and $\sim 10^{18}\,\rm g\,s^{-1}$ at the peak of the outburst. 
The quiescent accretion rate is two and a half orders of magnitude 
higher than predicted by the standard disc instability model, and we suggest 
this may be because the inner accretion disc in SS~Cyg is in a 
permanent outburst state. 

\end{abstract}


\section{Introduction}
\begin{figure*}
\begin{center}
\includegraphics{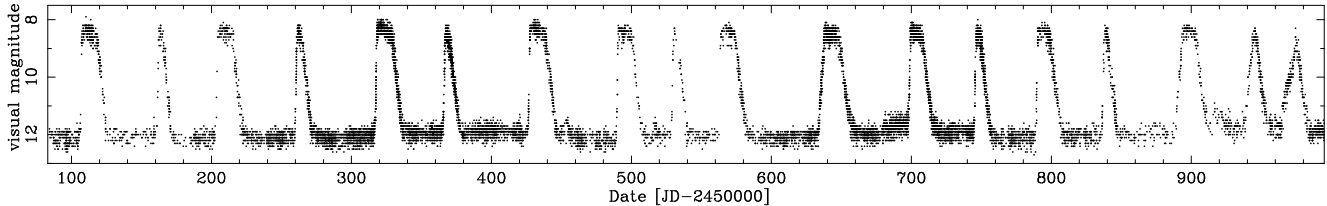}
\caption{\label{fig-lc} 
Two-and-a-half year portion of the AAVSO visual lightcurve of SS~Cygni.
The outburst discussed in this paper is centred on JD\,2\,450\,370.
}
\end{center}
\end{figure*}

Dwarf novae are remarkable binary stars that exhibit frequent large-amplitude
optical outbursts. They are relatively bright objects and their
optical lightcurves have been studied in detail by amateur astronomers
for more than a century. Figure\,\ref{fig-lc} shows a fragment of this coverage
of the brightest dwarf nova, SS~Cygni.

The outburst behaviour of dwarf novae has been interpreted as 
the result of 
a thermal-viscous instability
in the accretion disc surrounding
the white dwarf primary star \cite{Osaki74,Meyer81,Bath82}.
The disc instability model has been developed over many years
\egcite{Smak84,Cannizzo93a,Osaki96}
and though it has tended to lack predictive power, the 
model now accounts reasonably well for the observed
optical lightcurves. A recent review of its strengths and weaknesses
is given by \scite{Lasota01}.

Optical lightcurves of dwarf novae provide a reasonably good indicator of the 
behaviour of the outer accretion disc, which has temperatures in the range
$10^3$--$10^4$\,K. They tell us little, however, about the inner disc and
boundary layer where most of the gravitational energy is believed to be
released. The boundary layer is the region between the disc and the star where 
disc material loses kinetic energy and settles onto the white dwarf
surface. Temperatures of $10^5$-$10^8$\,K are generated 
in this region, so it must be studied in the extreme-ultraviolet and 
X-ray wavebands. 
Observations in these wavebands allow us to investigate the
energetics of dwarf nova outbursts and, in principle, can provide a
clean probe of the mass transfer rate through the boundary layer onto
the white dwarf. 
This is a key measurement for comparison with accretion disc models. 

Unfortunately, high-energy observations are rare because the
unpredictable nature of dwarf nova outbursts make them
difficult targets for space-based observatories, which tend
to be scheduled weeks or months in advance. The duration of dwarf nova
outbursts is also a problem because, lasting typically for a week,
they are too long to be covered by a normal pointed observation and
too short for repeated monitoring observations to be efficient. 

Some of the best high-energy coverage of dwarf nova outbursts
to date is that obtained with the
early scanning X-ray experiments. Despite poor signal-to-noise ratio, these
instruments scanned the same regions of sky many times over long
baselines and the resulting coverage was well suited to the outbursts
of dwarf novae. In particular, SS Cyg was well observed with 
Ariel V \cite{Ricketts79} and U~Gem with {\it HEAO}~1 \cite{Mason78,Swank78}. 
{\it EXOSAT\/} observations with higher signal-to-noise ratio 
but worse temporal 
coverage were made of SS~Cyg \cite{Jones92}, VW~Hyi \cite{Pringle87} and
U~Gem \cite{Mason88}. More recently, the {\it ROSAT\/} all-sky survey
provided the first combination of good sensitivity and temporal coverage,
resulting in relatively well-sampled outburst observations of 
SS~Cyg \cite{Ponman95},
VW~Hyi \cite{Wheatley96}, and 
Z~Cam \cite{Wheatley96b}. {\it ROSAT\/} was also used
to monitor the dwarf nova YZ~Cnc \cite{Verbunt99}.

{\it EXOSAT\/} and {\it ROSAT\/} 
had some sensitivity to the extreme-ultraviolet
waveband, but more recently the {\it Extreme Ultraviolet Explorer
(EUVE)} 
has provided the first detailed
view of dwarf nova outbursts in the extreme-ultraviolet. Rapid-response
target-of-opportunity observations allowed \euve\ to observe the
earliest moments of the extreme-ultraviolet outburst. Observations
were made of SS~Cyg \cite{Mauche95}, U~Gem \cite{Long96},  VW~Hyi
\cite{Mauche96} and OY~Car \cite{Mauche00}. 
\scite{Mauche01b} discuss all these observations and find
that extreme-ultraviolet flux rises dramatically 
0.7--1.5\,d
after
the initial optical rise. It peaks with or slightly after the optical,
declines gradually for a few days, and then declines more steeply than
the optical emission. 

While the extreme-ultraviolet behaviour of dwarf nova outbursts is now
well documented, the X-ray picture remains fragmentary and the
individual observations listed above 
yield only a
sketchy picture of the X-ray behaviour of dwarf novae through
outburst. The overall picture that emerges is that hard X-rays are
present in quiescence, but strongly suppressed during outburst (with
the exception of U~Gem; \ncite{Swank78}). The
X-ray suppression is approximately coincident with the sharp rise in the
extreme-ultraviolet, and the X-ray flux remains
low (but non-zero) throughout outburst until it recovers suddenly at
the very end of the optical outburst. 

\scite{Pringle79} and \scite{Patterson85} suggest the transition from 
X-ray to extreme-ultraviolet emission
occurs when the rising accretion rate causes
the boundary layer to become optically-thick to its own radiation. 
It then cools efficiently and the optically-thin hard X-ray emission
(kT$\sim$10\,keV) is replaced by intense optically-thick
extreme-ultraviolet emission (kT$\sim$10\,eV). 

This overall picture is consistent with most of the existing 
observations, but the various individual features have never been 
observed together in a single dwarf nova outburst. 
In this paper we present 
very long observations, in both the
X-ray and extreme-ultraviolet wavebands,  covering a full dwarf nova
outburst of SS~Cyg. 
We employ simultaneous target-of-opportunity observations with \euve\ 
and the {\it Rossi X-ray Timing Explorer (RXTE)}, triggered by amateur
astronomers, to provide the most complete multiwavelength coverage of
any dwarf nova outburst. 

\section{Observations} 
\subsection{Optical}
The American Association of Variable Star Observers (AAVSO) routinely
monitor a large number of variable stars, mainly visually with small
telescopes. 
In May 1996 we issued a request in AAVSO Alert Notice 221 
\cite{Mattei96a} to monitor SS Cyg particularly intensely in
preparation for our pre-approved target-of-opportunity observations 
with \rxte\   and \euve. We issued further requests in June and September 
in AAVSO Alert Notices 222, and 229 \cite{Mattei96b,Mattei96c} and
finally on 1996 October 9 we were able to trigger our satellite observations.
At 03:45\,UT an AAVSO observer in California, 
Tom Burrows, warned us that SS~Cyg was probably on the rise at 
$\rm m_{vis}$=11.5 and this was confirmed at 06:45\,UT by an observer in
Hawaii, Bill Albrecht, at which point it had reached $\rm m_{vis}$=10.9. 
We triggered our \rxte\ and \euve\ observations at around 07:15\,UT. 

The AAVSO 
lightcurve of SS~Cyg \cite{Mattei-SS}
is an invaluable resource, 
with a continuous record of observations 
dating back to the discovery of the system in 1896
\cite{Pickering1896}.
Figure\,{\ref{fig-lc}} shows the AAVSO lightcurve of SS~Cyg for
2.5\,yrs around the time of our \rxte\ and \euve\
observations. 

The
outburst properties
of SS~Cyg are analysed and discussed by \scite{Cannizzo92} and 
\scite{Cannizzo98}. They find the outburst durations of SS~Cyg form a bimodal
distribution with peaks at one and two weeks.
The decay timescales of all outbursts are similar, around 
2.4\,d\,mag$^{-1}$, but rise timescales range from 
0.5--3.5\,d\,mag$^{-1}$.
The outburst covered by our X-ray and extreme-ultraviolet observations
is narrow with a fast rise. 

\subsection{X-ray}
\rxte\ observations of SS~Cyg began at 1996 October 9 16:47\,UT, less
than ten hours after our trigger. We observed continuously for two
days in an attempt to cover the expected transition from dominant X-ray
to extreme-ultraviolet emission, and for three days at the end of the
outburst in an attempt to cover the reverse transition. Total exposure
times were 118\,ks and 175\,ks respectively, 
with gaps due only to Earth occultations and 
passage of the spacecraft through the South Atlantic Anomaly. 

We use data from the proportional counter array (PCA), which consists
of five xenon-filled proportional counter units (PCUs) labeled 0 to
4. PCUs 3 and 4 are often switched off due to discharge problems and
so we use only PCUs 0--2 for the lightcurves presented here. All \rxte\
count rates in this paper are for three PCUs. For
hardness ratios we use all five PCUs where available in order to maximise
the signal-to-noise ratio. 
We extract data only from the top xenon layer of each PCU
and accumulate lightcurves in the energy range 2.3--15.2\,keV corresponding to
PCA pulse height analysis (PHA) channels 5--41. For hardness ratios we split
the lightcurves at PHA channel 15 (5.8\,keV).

For lightcurves we exclude data only if the elevation about the
Earth's limb is less than 3\,deg. For spectra we exclude data where
the elevation is less than 10\,deg, and also for 30\,min after
passage through the South Atlantic Anomaly and whenever the electron 
contamination is
$>$10\%. We background subtract our data using the faint source
(L7/240) background model.

\subsection{Extreme ultraviolet}
\begin{figure}
\begin{center}
\includegraphics[width=8.5cm]{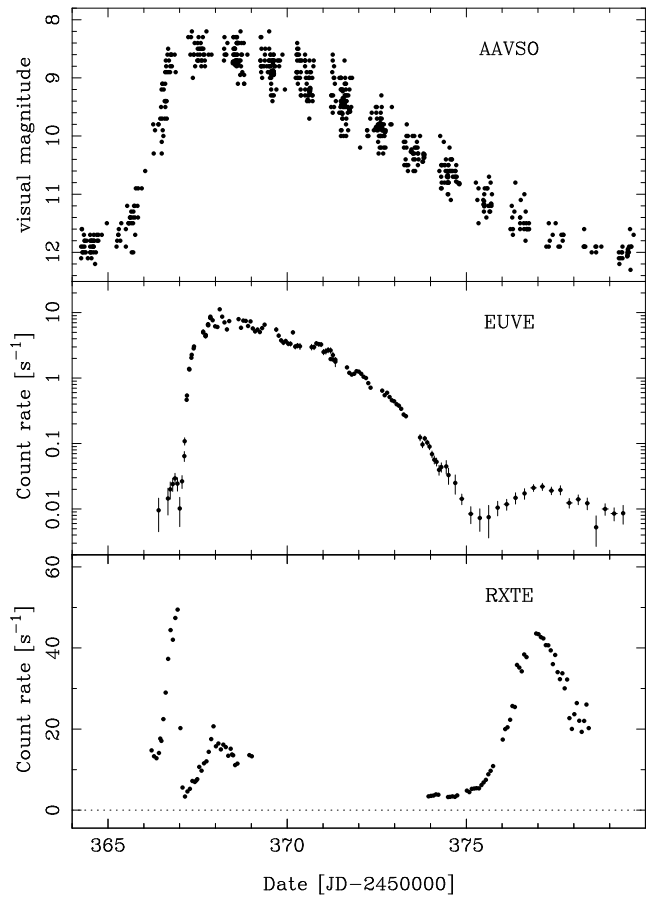}
\caption{\label{fig-all} 
Simultaneous AAVSO, \euve\ and \rxte\ observations of SS~Cygni
throughout outburst. The \euve\ lightcurve is a combination of DS and
SW lightcurves plotted on a log intensity scale (scaled to the DS
count rate). The AAVSO lightcurve is also plotted on a log scale (visual
magnitude), but the \rxte\ lightcurve is plotted on a linear scale in
order to emphasize the timing of the sharp transitions. 
}
\end{center}
\end{figure}

\euve\ \cite{Bowyer91,Bowyer94} observations of SS Cyg began on 1996 
October 9 21:39 UT and were performed continuously for 13 days. \euve\ data 
are acquired only during satellite night, which comes around every 95 
min and lasted during our observation for between 23 min and 32 min. Valid
data are collected during intervals when various satellite housekeeping 
monitors (including detector background and primbsch/deadtime corrections)
are within certain bounds. After discarding numerous short ($\Delta t\le 
10$~min) data intervals comprising less than 10\% of the total exposure, 
we were left with a net exposure of 208 ks. Unfortunately, the Deep
Survey (DS) photometer
was switched off between October 11 8:46 UT and October 14 16:53 UT because
the count rate was rising so rapidly on October 11 that the \euve\ 
Science Planner feared that the DS instrument would be damaged while 
the satellite was left unattended over the October 12--13 weekend. We
constructed an extreme-ultraviolet light curve of the outburst from the
background-subtracted 
count rates registered by the DS photometer and the Short Wavelength
(SW) spectrometer. We used a 
72--130
\AA \ 
wavelength cut for the SW spectrometer data
(minus the 76--80~\AA \ region, which is contaminated by scattered 
light from an off-axis source)
and applied an empirically-derived scale factor of 14.93 to the 
SW count rates to match the DS count rates. The resulting
extreme-ultraviolet light curve
is shown in the middle panel of Figure\,\ref{fig-all}. 
We typically plot one point per
valid data interval, although seven valid intervals between October 10 
15:13 UT and October 11 8:35 UT were split to better define the rise of
the extreme-ultraviolet 
light curve, and data after October 18 4:45 UT were binned to
0.25 days (3--4 satellite orbits) to increase the signal-to-noise ratio.
Photometric aspects of the \euve\ data have been reported previously by
\scite{Mauche01a}, and we refer the reader to that paper for
additional details.

\section{Time Series Analysis}
\subsection{Outburst lightcurves}
\label{sect-lcs}
Figure\,\ref{fig-all} shows the optical, 
{\it EUVE\/}, and {\it RXTE\/} light curves
of SS~Cyg 
throughout outburst.
The early alert from the
AAVSO and the timely response of the satellite schedulers allowed us to
get on target exceptionally early: within a day of the beginning of
the optical rise at JD\,2\,450\,365.0--2\,450\,365.5. 
As is typical of fast-rise outbursts of SS~Cyg,
the optical light curve rose gradually
at first and then rapidly, 
crossing $\rm m_{vis}$=11 at about JD\,2\,450\,365.9, and
reaching maximum ($\rm m_{vis}$$\sim$8.5) about a day later 
(the uneven spacing of the optical observations limit the precision of
these time estimates).
SS~Cyg remained at maximum optical light for at most one
day, then declined gradually 
for a few
days ($\sim$0.08\,mag\,d$^{-1}$), and then more rapidly 
($\sim$0.45\,mag\,d$^{-1}$), returning to minimum on about JD\,2\,450\,378.

The rise at 
high energies
began in the hard X-ray band at
JD\,2\,450\,366.4, about 0.9--1.4\,d after the beginning of the rise in the
optical. A rise in the hard X-ray flux at the beginning of an outburst 
has been previously suspected \cite{Swank79,Ricketts79}, but
this is the first time it has been observed unambiguously. 
The {\it RXTE\/} count rate rose by a factor of 4 over half a day, 
but was then suddenly quenched to near zero in less than 3\,h. 
This sudden suppression of the hard X-ray flux was followed
immediately by a rapid rise in the extreme-ultraviolet
flux. The transition from X-ray to extreme-ultraviolet 
emission has previously been
inferred from fragmentary data \egcite{Jones92}, but it is
resolved here for the first time. The temporal coincidence of the hard
X-ray suppression and the extreme-ultraviolet rise leaves no doubt that both
components arise from the same source, 
presumably the boundary layer between the accretion disc and the 
surface of the white dwarf (see also Sect.\,\ref{sect-dis_bl}).

The dramatic rise of the \euve\ lightcurve began at JD\,2\,450\,367.0, about
1.5--2\,d after the beginning of the rise in the optical, and 0.6\,d
after the beginning of the rise in X-rays. 
In contrast to the optical lightcurve, the rise of the extreme-ultraviolet 
lightcurve  was rapid at first, then gradual: 
the {\it EUVE\/} count rate rose by
two orders of magnitude in the first 0.5\,d, then by a factor of 3 in
the next 0.5\,d. 
After reaching maximum at JD\,2\,450\,368.0, the \euve\ lightcurve began
to decline almost immediately, first gradually and then more rapidly. 
The decline of the {\it EUVE\/} count rate is nearly exponential,
with an e-folding time 2.3\,d
during the first few days of the decline, steepening to 
0.52\,d
during the last few days of the decline, and reaching minimum at about 
JD\,2\,450\,375.5.
After that time, both the extreme-ultraviolet and hard X-ray count
rates increased again to a peak at about JD\,2\,450\,377.0 
before declining again, presumably to their quiescent levels. 

The extreme-ultraviolet lightcurve at the very beginning (before
JD\,2\,450\,367) and end (after JD\,2\,450\,376) of the observation
exactly follows the hard X-ray lightcurve, 
suggesting that the photons detected by {\it EUVE\/} at those times 
are dominated by the soft tail of the optically-thin, 
hard X-ray emission component. This reminds us that the 
{\it EUVE\/} and {\it RXTE\/} count rate lightcurves are driven in general 
by both flux and spectral variations.

\subsection{Hardness ratios}
\label{sect-hard}

\begin{figure}
\begin{center}
\includegraphics{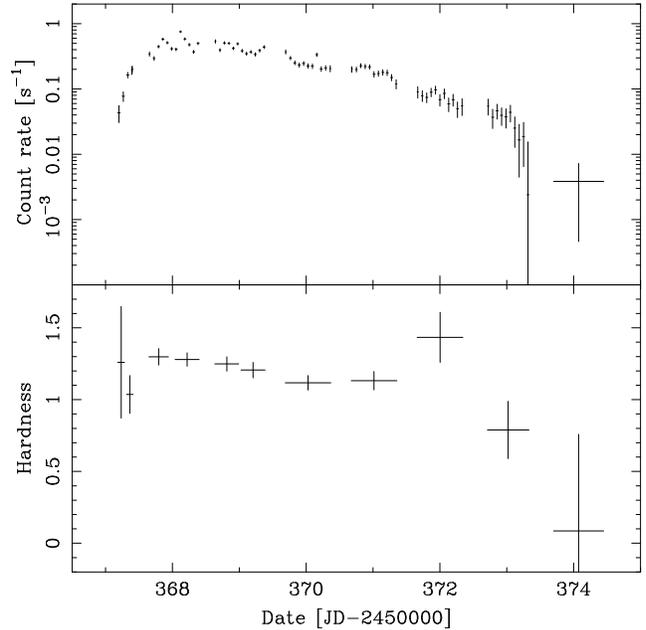}
\caption{\label{fig-euve_hard}
The \euve\ SW lightcurve and hardness ratio of SS~Cyg in outburst. 
The hardness ratio is defined as {\it H/S} where 
{\it S} is the count rate in the 95--130\,\AA\ band and
{\it H} is the count rate in the 
72--95\,\AA\ band, minus the 
76--80\,\AA\ band which is
contaminated by scattered light from an off-axis source.
}
\end{center}
\end{figure}

\begin{figure}
\begin{center}
\includegraphics{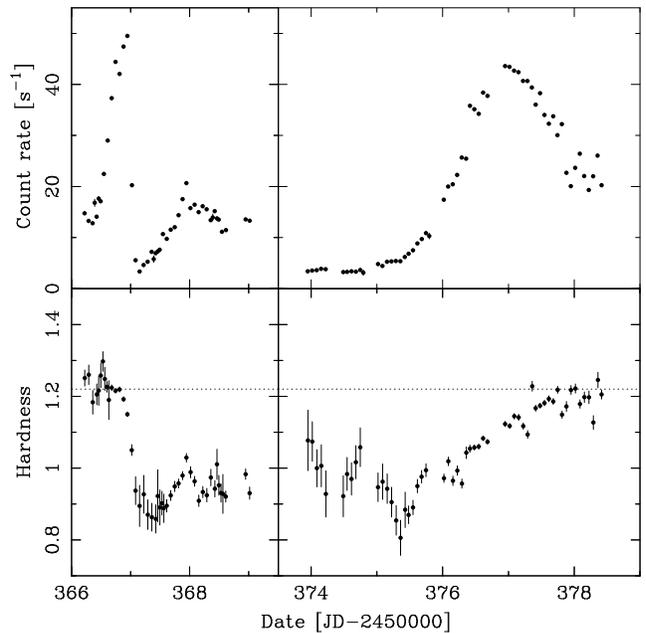}
\caption{\label{fig-xte_hard}
The \rxte\ 2--15\,keV lightcurve and hardness ratio of SS~Cyg in outburst. 
The hardness ratio is defined as {\it H/S} where {\it H} is the count rate 
in the 6--15\,keV band and {\it S} is the count rate in the 2--6\,keV band.
}
\end{center}
\end{figure}

\begin{figure*}
\begin{center}
\includegraphics{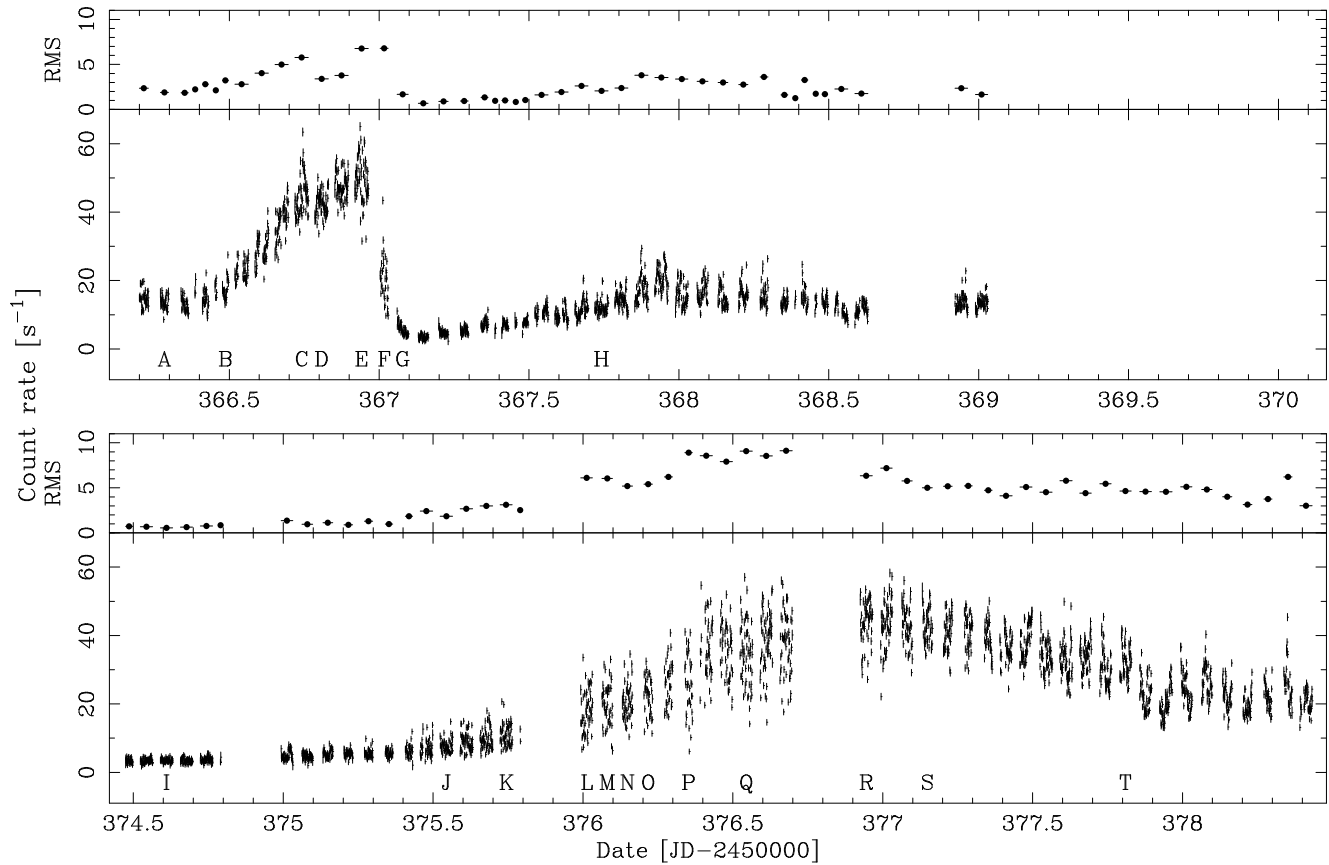}
\caption{\label{fig-var}
The \rxte\ 2--15\,keV lightcurve of SS~Cyg binned at 64\,s. 
The top panels show the observations at the beginning of outburst; the
bottom panels show those at the end of outburst.
The smaller panels show the RMS variability measured within each observing 
interval.  
It can be seen that the X-ray suppression and recovery are accompanied
by intense variability. 
The letters correspond to intervals of the 16\,s-binned lightcurves in
Figs.\,\ref{fig-detail}\,\&\,\ref{fig-hardness_detail}.
}
\end{center}
\end{figure*}

\begin{figure}
\begin{center}
\includegraphics{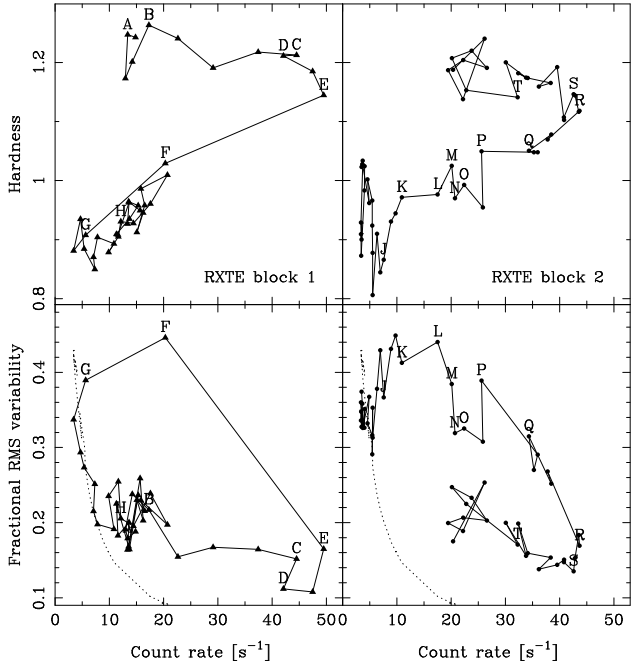}
\caption{\label{fig-xte_cr}
The variation of RXTE hardness ratio (top) and fractional variability 
amplitude (bottom) plotted against RXTE count rate. 
It can be seen that the first block of RXTE data (left) 
and the second block (right) follow the same paths but in opposite 
directions. 
Labels correspond to those defined in Fig.\,\ref{fig-var}. 
The dotted lines in the lower panels indicated the variability expected 
due to counting statistics alone. 
}
\end{center}
\end{figure}

To investigate the spectral variations in the extreme-ultraviolet 
and hard X-ray flux of
SS~Cyg, we divided the \euve\ SW and \rxte\ PCA counts into
soft ($S$) and hard ($H$) spectral bands, and formed the hardness ratio
$H/S$ for each. For the \euve\ SW we follow 
\scite{Mauche95}
and defined 
$S$ to be the count rate in the 95--130~\AA \ band, 
and $H$ to be the count rate in the 72--95~\AA \ band 
(but excluding the 76--80~\AA \ band, which is
contaminated by scattered light from an off-axis source).
For the {\it RXTE\/} PCA, $S$ is the count rate in the
2.3--5.8 keV band and $H$ is the count rate in the 5.8--15.2 keV band.
In each case, the bandpasses were chosen to produce roughly equal
number of counts in the two bands.

\subsubsection{Extreme-ultraviolet hardness ratio}

Figure 3 shows the {\it EUVE\/} SW count rate ({\it H+S}) 
and hardness ratio ({\it H/S}) as a function of time. 
Unfortunately, because of the small effective
area and the effectively high background of the SW spectrometer, useful
SW data exists for only the bright portion of the outburst between 
JD\,2\,450\,337.2--2\,450\,374.4. 
The figure shows that there is in general a correlation between the 
SW count rate and hardness ratio, with hardness ratios of 1.0
on the rise to outburst, 1.3 at the peak of the outburst,
a gradual decline to 1.1, a possible increase to 1.5, 
and then a more rapid decline to 0 at the end of the
interval. 
Although the signal-to-noise ratio is low, 
a constant hardness ratio is ruled out at the 97\% confidence level
($\chi^2$ of 20.7 with 10 degrees of freedom)
and a linear fit is only marginally acceptable 
($\chi^2$ of 13.1 with 9 d.o.f.). 

The observed range of hardness ratio variations is far broader than that 
seen by
\scite{Mauche95}
during the rise of the slow-rise 1993 August outburst
of SS~Cyg, where the hardness ratio remained at 1.4 (1.7 after
similarly omitting the 76--80\,\AA\ band) during an increase of two 
orders of magnitude in the SW count rate. 

As discussed by 
\scite{Mauche95}
and in Sect.\,\ref{sect-euve_spec}, SW hardness ratio
variations can be due to changes in the effective temperature of 
the boundary layer spectrum and/or changes in the
absorbing column density. The higher hardness ratios near the peak of the
outburst imply that at those times the boundary
layer emission of SS~Cyg is hotter and/or more strongly absorbed. 

An
increase in the boundary layer temperature during outburst is expected
from the blackbody relation $T=(L_{\rm bb}/4\pi\sigma f R_{\rm wd}
^2)^{1/4}$, where $L_{\rm bb}\approx GM_{\rm wd}\dot{M}/2R_{\rm wd}$
is the blackbody luminosity, $M_{\rm wd}$ and $R_{\rm wd}$ are the white
dwarf mass and radius respectively, $\dot{M}$ is the mass-accretion rate,
and $f$ is the fractional emitting area: assuming that $f$ remains
constant, $T\propto \dot{M}^{1/4}$. An increase in the effective
absorbing column density is expected if the boundary layer is viewed
through the high-velocity wind driven off the accretion disc during
outburst. 

Whatever the cause, Fig.\,\ref{fig-euve_hard} 
demonstrates that the {\it EUVE\/} count rate evolution during the
outburst is driven to some degree by changes in the shape of the
spectrum. We provide more details in 
Sect.\,\ref{sect-euve_spec}.

\subsubsection{X-ray hardness ratio}
\label{sect-hard_xte}

Figure\,\ref{fig-xte_hard} shows the \rxte\ X-ray hardness ratio of
SS~Cyg through the outburst. It can be seen that the sharp X-ray 
suppression at the beginning of outburst is accompanied by a rapid 
softening of the X-ray spectrum from {\it H/S}=1.2 to 0.9. 
The softening begins around two \rxte\ orbits before the rapid
count-rate suppression:  at the same time as the \rxte\ rise begins to slow. 

The X-ray suppression is followed by a
secondary X-ray peak that follows the extreme-ultraviolet brightness. 
During this secondary peak there is a corresponding re-hardening of the
X-ray spectrum. 

In the second block of \rxte\ observations, the hardness ratio first
seems to drop from 1.0 to 0.8 before returning to the 
quiescent value of 1.2. The return to the
quiescent value occurs much more slowly than at the time of the X-ray
suppression, proceeding gradually throughout the X-ray recovery, peak
and subsequent decline to quiescence. 

The timing of the sudden X-ray softening in the first block of \rxte\ 
observations is clearly associated with
the suppression of the X-ray emission, presumably occurring as the 
boundary layer becomes optically thick to its own radiation. 
The gradual hardening at the end of the
outburst (in the second block of observations) 
shows that the reverse transition occurs more slowly,
presumably because the accretion rate is changing more slowly on
decline. 

It seems then that the X-ray hardness correlates with the accretion
rate during
the boundary layer transitions. It may be the best indicator of
accretion rate at these times, as the X-ray and extreme-ultraviolet
count rate lightcurves are probably both dominated by the changing
spectrum. These spectral variations are investigated in more detail in 
Sect.\,\ref{sect-xte_spec} and by Wheatley (in preparation).

The symmetry of the X-ray transitions at the beginning and end of outburst 
is underlined by the top panels of Fig.\,\ref{fig-xte_cr} where we plot 
the hardness ratio against count rate for the two blocks of RXTE observations. 
The labels A--T correspond to those in Fig.\,\ref{fig-var}.
Despite the differing timescales, it can be seen that the two blocks 
follow the same path in hardness-intensity space (but in opposite directions). 
During the X-ray suppression (intervals E--G), outburst (G--K) and 
recovery (K--R), the X-ray hardness ratio is strongly 
correlated with count rate. In early outburst (A--E) and late outburst 
(after interval R) the hardness ratio is weakly anti-correlated with the 
count rate. 

\subsection{Rapid X-ray variability}
\label{sect-var}

\begin{figure*}
\begin{center}
\includegraphics{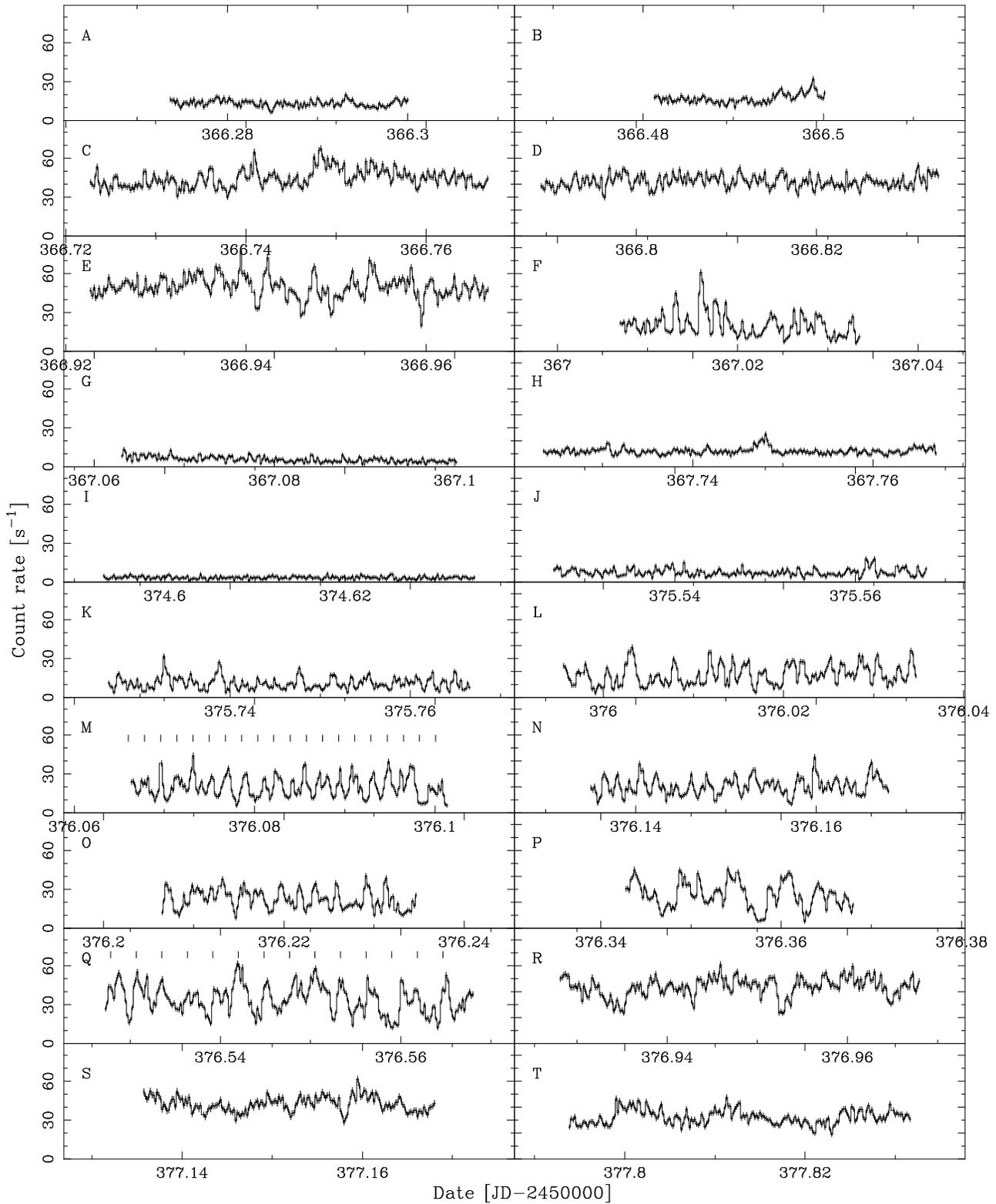}
\caption{\label{fig-detail}
\rxte\ 2--15\,keV lightcurves of SS~Cyg binned at 16\,s, 
with labels corresponding to those in Fig.\,\ref{fig-var}. 
The intense variability at X-ray suppression and recovery is clearly
resolved and appears to take the form of high-amplitude
quasi-periodic oscillations. The tick marks in panels M and Q
correspond to the characteristic periods found in
Sect.\,\ref{sect-var} and plotted in Fig.\,\ref{fig-lombslots}.
}
\end{center}
\end{figure*}

Figure\,\ref{fig-var} shows the \rxte\ lightcurve rebinned at 64\,s. 
It
is clear that there are large-amplitude rapid variations in addition to
the overall flux evolution. The variations are strongest at the time
of the hard X-ray suppression around JD\,2\,450\,367.0, and at the
time of the X-ray recovery on JD\,2\,450\,376.
It is particularly striking that the variations during the recovery are
much stronger than those seen at the same mean count rate during the
subsequent decline to quiescence on JD\,2\,450\,377.

The periods of strong variability coincide with the changes in the
X-ray hardness ratio and are presumably 
associated with the transition of the boundary layer between its 
optically-thin and optically-thick states. 
It seems these transitions do not occur smoothly. 

The bottom panels of Fig.\,\ref{fig-xte_cr} show the fractional variability 
amplitude in each of the intervals of the 16\,s binned lightcurve plotted 
against count rate. The labels A--T correspond to those in 
Fig.\,\ref{fig-var}.
Once again, the two blocks of the RXTE data follow the 
same paths but in opposite directions. It can be seen that the 
variability amplitude peaks at 45 per cent during the X-ray suppression 
(interval F) and at the beginning of the X-ray recovery (intervals K and L). 
During the two X-ray transitions (intervals E--F and K--S) the variability 
decreases with count rate to a minimum of around 12 per cent. 
At all other times, 
i.e. before the X-ray suppression, during the outburst and after the X-ray 
recovery, the variability is weakly anti-correlated with count rate, 
ranging between 15 and 20 per cent. At times of very low count rate 
(less than about 10\,$\rm s^{-1}$) the variability is dominated by counting 
statistics. 

Figure\,\ref{fig-detail} shows a close-up view of the variability 
during various intervals in the \rxte\ lightcurve, which has been 
binned at 16\,s.
The variability during X-ray suppression and recovery seems to take the form of
sharp dips and peaks with in some cases factors of two or more
variability in less than a 16\,s bin. At times these dips and peaks
seem to take the form of quasi-periodic oscillations, perhaps
oscillations in the state of the boundary layer. 

\begin{figure}
\begin{center}
\includegraphics{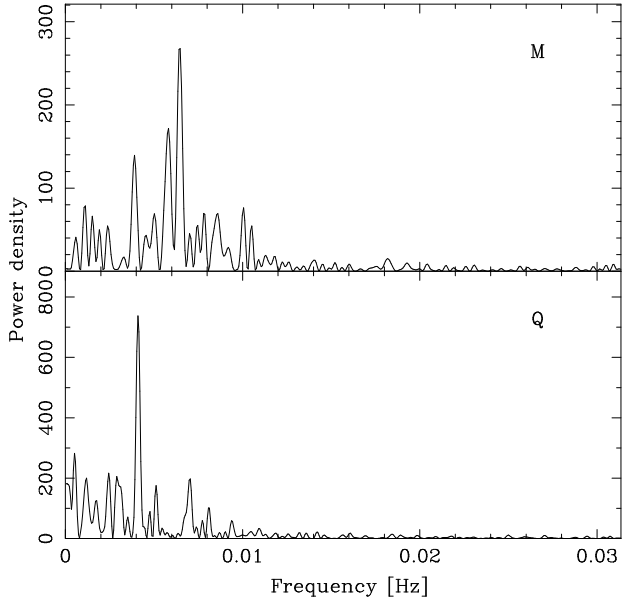}
\caption{\label{fig-lombslots}
Power spectra of two portions of the \rxte\ lightcurve of SS~Cyg. The
labels M and Q refer to the intervals labeled in 
Figs.\,\ref{fig-var}\,\&\,\ref{fig-detail}. This is described and
discussed in Sect.\,\ref{sect-var}.
}
\end{center}
\end{figure}

We calculated power spectra for each lightcurve interval using the 
Lomb-Scargle algorithm as implemented by \scite{NumRec2} 
and with a slight modification of the normalisation of the power
spectrum (using the expected variance rather than the measured variance).
These spectra show that the oscillations apparent in Fig.\,\ref{fig-detail}
are not strictly periodic. They do, however, allow us to identify
characteristic periods, for instance, 
155\,s in interval M and 245\,s in interval Q. 
The power spectra for these two intervals are plotted in 
Fig.\,\ref{fig-lombslots} and tick marks corresponding to these
periods have been included in Fig.\,\ref{fig-detail}.
The imperfect match between peaks and tick marks 
emphasises the quasi-periodic nature of the oscillations. 

\begin{figure}
\begin{center}
\includegraphics{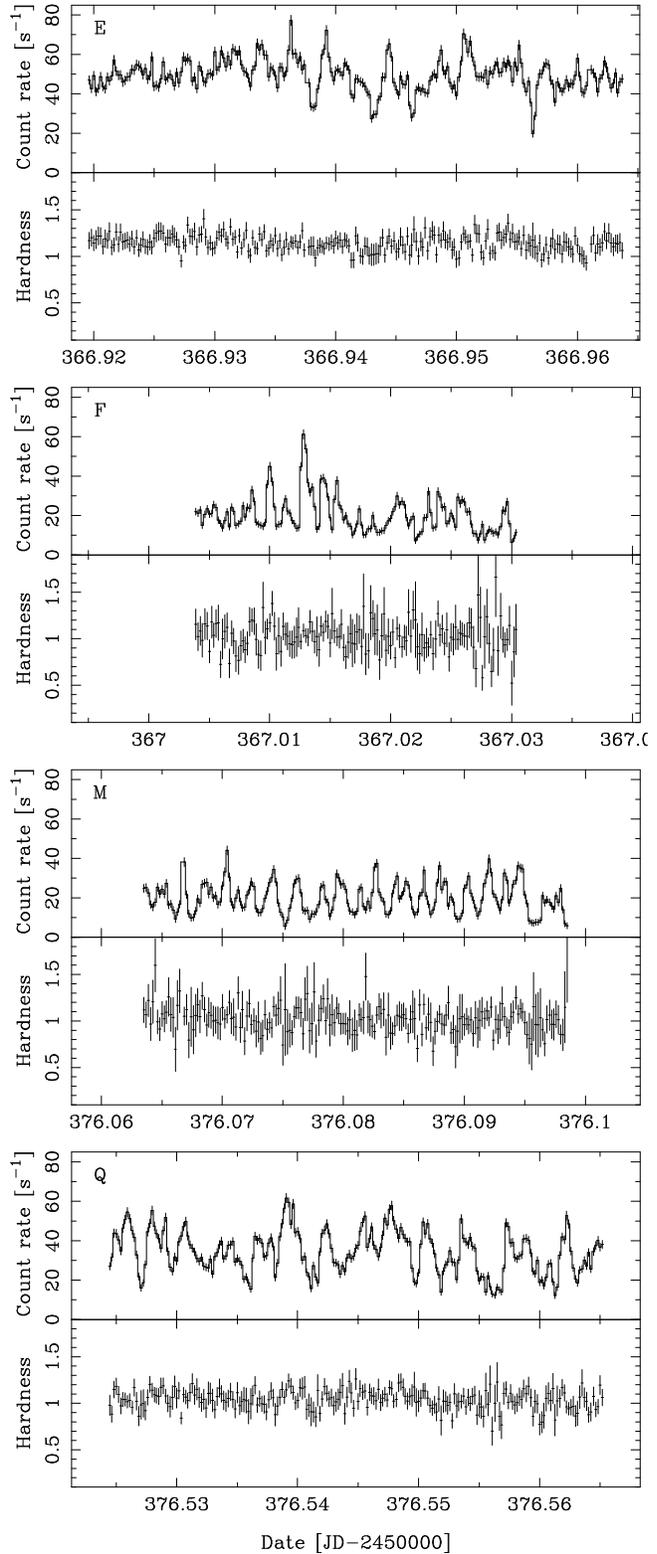}
\caption{\label{fig-hardness_detail}
The \rxte\ 2--15\,keV lightcurve and hardness ratio of SS~Cyg at the times of
X-ray suppression and recovery. The labels correspond to those in
Figs.\,\ref{fig-var}\,\&\,\ref{fig-detail}. 
The lightcurve and hardness ratio are binned at 16\,s. Remarkably
the rapid X-ray variability is not associated with hardness variations. 
}
\end{center}
\end{figure}

\begin{figure}
\begin{center}
\includegraphics{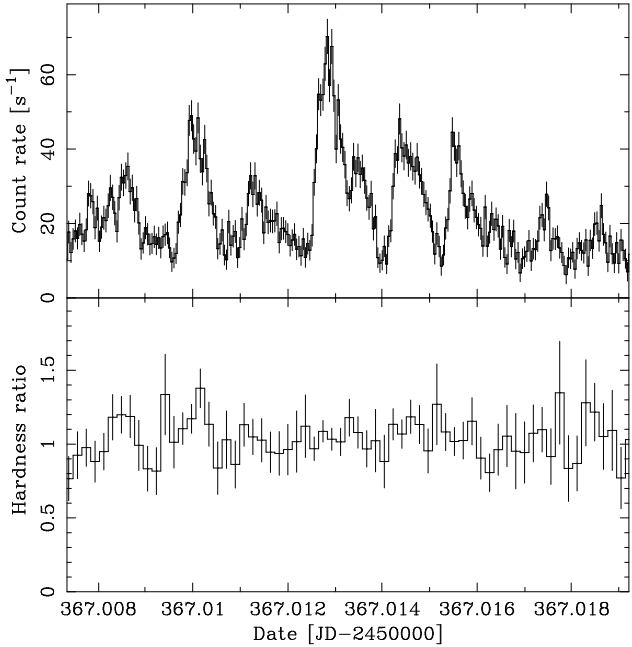}
\caption{\label{fig-slot15}
Close-up of the \rxte\ 4\,s-binned lightcurve at the time of the X-ray
suppression (interval F in Fig.\,\ref{fig-var}). 
The lack of hardness ratio variations rule out temperature or absorption 
column density changes as the source of the large peaks. In
Sect.\,\ref{sect-cool} we use the resulting limit on the cooling timescale
to measure a lower limit to the density of the X-ray emitting gas.
}
\end{center}
\end{figure}

Remarkably, Figs.\,\ref{fig-hardness_detail}\,\&\,\ref{fig-slot15}
show that these strong variations/oscillations are not associated
with any significant hardness variations. 
This rules out temperature changes and photoelectric absorption as the 
source of the variability (although occultation by sufficiently 
optically-thick ``bricks'' cannot be ruled out). 
Instead the rapid X-ray variability must be due to changes in the emission 
measure of the X-ray emitting plasma: either oscillations in the density of 
the emitting region or changes in the accretion rate through the 
optically-thin portion of the boundary layer. 

\subsubsection{X-ray cooling time-scale and gas density}
\label{sect-cool}
Resolving X-ray variability allows us to constrain the
density of the X-ray emitting plasma. This is because the flux cannot
drop faster than the cooling timescale of the gas, which is a strong
function of density. 

X-ray emission from dwarf novae is thought to arise in a cooling flow, 
in which we see gas in pressure equilibrium cooling from
a post-shock maximum temperature all the way down to the white dwarf 
photosphere temperature \egcite{Wheatley96,Done97,Mukai03}. 
A sudden change in accretion rate results in an adjustment of the cooling flow
on the timescale of the slowest cooling gas. As long as the shock
temperature remains unchanged the cooling flow spectrum will not
change, but the emitted flux will increase or decrease to the new
equilibrium value. 

Figure\,\ref{fig-slot15} shows a close-up view of the \rxte\  
lightcurve with 4\,s bins at the time of the X-ray suppression
(interval F in 
Figs.\,\ref{fig-var},\,\ref{fig-detail}\,\&\,\ref{fig-hardness_detail}). 
There are a series of large amplitude peaks in the X-ray emission, none of 
which are accompanied by changes in the hardness ratio. 
The cooling timescale of the X-ray plasma 
must be shorter than the decline from these peaks, 
allowing us to place a limits on the density of the hottest plasma
(which is the slowest to cool). 

Our spectral analysis (Sect\,\ref{sect-xte_spec}) shows that the
hard X-rays detected by \rxte\  are dominated by emission at temperatures
around 10\,keV. At such temperatures the dominant cooling
mechanism is bremsstrahlung radiation which is emitted proportionally
with the square of the electron density. 

The thermal energy of the X-ray emitting gas, E, is given by 
$E = \frac{3}{2} k T n_e V\rm \,erg$ and its bremsstrahlung luminosity, 
$L$, 
by $L= 1.64\times10^{-27} T^{1/2} g(T) n_e^2 V\rm\,erg\,s^{-1}$, where 
$k$ is the Boltzmann constant, $T$ is temperature of the gas, $n_e$ is
the electron density, $V$ is the volume, and $g(T)$ is the Gaunt factor. 
Taking the ratio $E/L$ gives a characteristic time-scale for the cooling
of the X-ray emitting gas, 
$\tau_{\rm cool}=1.3\times10^{11} T^{1/2} g(T)^{-1} n_e^{-1}\rm\,s$. 
For a given temperature this is simply inversely proportional to the
electron density. 
For a 10\,keV gas
$\tau_{cool} n_e=1.0\times10^{15}\rm\,s\,cm^{-3}$. 

Inspecting the lightcurve in 
Fig.\,\ref{fig-slot15} we see the X-ray flux can drop by a factor two in 
20\,s without variations in the hardness ratio. 
The cooling timescale of the X-ray emitting plasma must then be 
less than 20\,sec, limiting the electron number density to 
$n_e\geq5\times10^{13}\rm\,cm^{-3}$. 

Taking the appropriate emission measure from our spectral fitting 
(Sect.\,\ref{sect-xte_spec}; 
$8\times10^{54}\,d_{\rm 100pc}^2\rm\,cm^{-3}$) 
this gives an upper limit to the emitting volume of the $\sim$10\,keV
plasma of
$V\leq3\times10^{27}\rm\,cm^3$. For comparison the volume of the white 
dwarf in SS~Cyg is approximately $7\times 10^{26}\rm\,cm^3$ 
($\rm R_{wd}\sim5.5\times10^{8}\,cm$).

\subsection{Dwarf nova oscillations}
\label{sect-dnos}
Dwarf nova oscillations are detected in our \euve\ lightcurve of
SS~Cyg and are discussed in detail by \scite{Mauche01a}. They are
first detected on the rise of the extreme-ultraviolet outburst and are
detected well into the fast-decline phase. The oscillation
period is anti-correlated with
the \euve\ DS count rate dropping from 7.81\,s to 6.59\,s on the rise,
before jumping to 2.91\,s and continuing to drop to 2.85\,s when the
DS was turned off. When the DS was turned back on the period was seen
to increase again on the decline from 6.73\,s to 8.23\,s. 

Since extreme-ultraviolet oscillations are seen on the rise, we might
expect to see corresponding oscillations in the \rxte\ X-ray lightcurve,
particularly on the initial X-ray rise and perhaps in the secondary
peak after the X-ray suppression. 

\begin{figure}
\begin{center}
\includegraphics{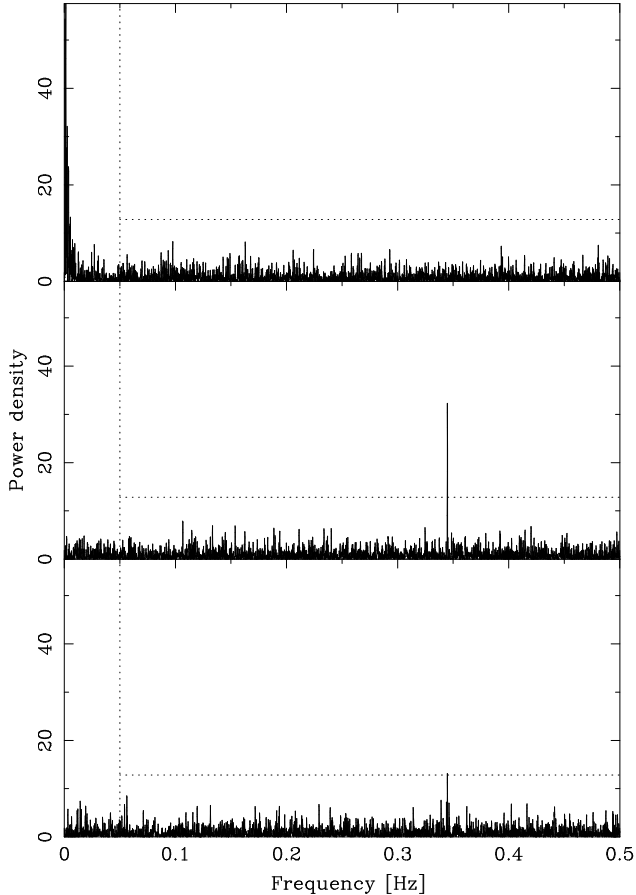}
\caption{\label{fig-slot28}
Top: power spectrum of interval H of the \rxte\ lightcurve of SS~Cyg
(see also Figs.\,\ref{fig-var}\,\&\,\ref{fig-detail}). No dwarf nova
oscillations are detected. The horizontal dotted line represents the 99\% 
confidence limits for detection of a period in the presence of white 
noise only. The vertical line represents the limit of the frequency range 
searched for periodic modulations.  
Middle and bottom: power spectra of simulated lightcurves of interval H 
containing 2.9\,s sine modulations of 7\% and 5\% amplitude respectively. 
See text for details of simulations. 
}
\end{center}
\end{figure}

In order to search for X-ray oscillations we divided our \rxte\
data into 103 individual lightcurves, binned at 1\,s, and each
with no data gaps longer than 1.5\,ks. We calculated the power
spectrum of each lightcurve using the Lomb-Scargle algorithm as
in Sect.\,\ref{sect-var}. As an example, the top panel of 
Figure\,\ref{fig-slot28} shows the power spectrum of the 1\,s lightcurve of 
interval H (labeled in Figs.\,\ref{fig-var}\,\&\,\ref{fig-detail}). 
This interval coincides with the detection of dwarf nova oscillations 
near 2.9\,s with \euve\ \cite{Mauche01a}.  

Low frequencies in our power spectra are dominated by the presence of strong
and non-stationary red noise. We therefore search for dwarf nova oscillations 
only in the period range 2--20\,s where individual inspection shows the power 
spectra to be dominated by white noise
(e.g.\   the top panel of Fig.\,\ref{fig-slot28}).
We tested the significance of peaks in this range using the method described 
in \scite{NumRec2}. The 99\% confidence limit for detection of period signals 
is plotted as a horizontal dotted line in Fig.\,\ref{fig-slot28}. 

No significant periodicities are found in any of our 103 \rxte\ lightcurves in 
the period range 2-20\,s. In four power spectra we do find peaks that exceed 
our significance threshold, but in all four cases the selected periods are at 
the low-frequency end of our search range (18.6, 18.9, 18.8 \& 15.7\,s) 
and we attribute these to red noise. The periods detected with \euve\ are in 
the range 2.8--8.2\,s, and as the \rxte\ count rate in the initial 
X-ray peak is more than an order of magnitude higher than that detected in 
the \euve\ DS instrument, we would certainly have expected to detect 
the equivalent oscillations in the X-ray band if present. 

In order to define an upper limit to such oscillations in the \rxte\
lightcurve we simulated lightcurves with the same duration, time
resolution and statistical quality as the \rxte\ interval H, 
but with a 2.9\,s sine modulation added. We simulated
lightcurves for sine amplitudes of 5\%, 7\% and 10\% of the source
count rate, 12\,s$^{-1}$. In each case we simulated five lightcurves 
(fifteen in all). The lower panels of Fig.\,\ref{fig-slot28} show 
example
power spectra for the 7\% and 5\% cases. 
The 7\% modulation was clearly detected in all five simulations. 
The 5\% modulation was detected in three of the five simulations. 
We thus take 7\% to be a conservative upper limit to the amplitude of 
dwarf nova oscillations in the hard X-ray emission of SS~Cyg. 
The observed amplitude in the \euve\ lightcurve around this time is 25--30\%. 

The absence of oscillations in the X-ray lightcurve suggests that dwarf
nova oscillations are a property only of the optically-thick
thermally-collapsed boundary layer. Whatever mechanism drives
dwarf nova oscillations, it does not seem to operate in the
optically-thin boundary layer prior to the the X-ray suppression, or 
in the source of the residual X-rays emitted during outburst 
(which may arise from a site other than the boundary layer, 
see Sect.\,\ref{sect-dis_xob}).

A second interpretation is that the oscillations are present in the
heating of the X-ray gas but that its cooling timescale is
sufficiently long to smooth out the oscillations. Using the timescale 
calculated in Sect.\,\ref{sect-cool} we find an electron number
density of $\rm n_e \leq 1\times10^{14}\,cm^{-3}$ would 
hide the slowest observed \euve\ oscillation period of 8.23\,s. 
As variations only a little longer than this are readily apparent in the 
X-ray lightcurve (Sect.\,\ref{sect-cool}; Fig.\,\ref{fig-hardness_detail}), 
we regard this interpretation as unlikely. 

\section{Spectral analysis}
To gain an understanding of the spectra of the hard and soft components
of the X-ray spectrum of SS~Cyg, to move beyond the hardness ratio
variations discussed in 
Sect.\,\ref{sect-hard},
we studied the {\it EUVE\/} SW and {\it RXTE\/} PCA spectra.
The PCA spectra are discussed in more detail by Wheatley (in preparation).

\subsection{Extreme-ultraviolet spectra}
\label{sect-euve_spec}
The mean extreme-ultraviolet spectrum of SS~Cyg in outburst 
was derived from {\it EUVE\/}
SW event data collected between JD\,2\,450\,367.2 and 2\,450\,372.5. After
discarding numerous short ($\Delta t<$10\,min)
data intervals
comprising less than 10\% of the total exposure, we were left with a
deadtime and primbsch-corrected exposure of 75.1\,ks. As before, we
excluded events between 76--80\,\AA \ to avoid contamination by scattered
light from an off-axis source, and binned the counts to $\Delta\lambda$
=0.2\,\AA , roughly half the FWHM (=0.5\,\AA ) spectral resolution of
the spectrometer \cite{Abbott96}.

\begin{figure}
\begin{center}
\includegraphics[width=8.5cm]{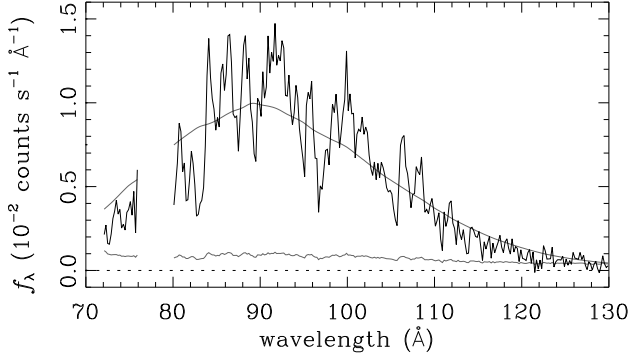}
\caption{\label{fig-euve_spec}
\euve\ SW count spectrum accumulated between JD\,2450367.2 and
2450372.5. The trace at the bottom of the figure is the $1\, \sigma $ error
vector and the smooth curve is an absorbed blackbody model. Exposure time
is 75.1 ks, the bin width $\Delta\lambda=0.2$~\AA , and the observed
SW count rate is 0.28~\rm counts~s$^{-1}$.
}
\end{center}
\end{figure}

The resulting spectrum is shown
in Fig.\,\ref{fig-euve_spec}, 
where the jagged upper curve is the background-subtracted
count spectrum and the lower curve is the associated $1\, \sigma $ error
vector. Comparing that figure with Fig.\,6 of 
\scite{Mauche95}
reveals that
the extreme-ultraviolet 
spectrum of SS~Cyg differs little from one outburst (or one
{\it type\/} of outburst) to the next. As before, the spectrum appears
to consist of a continuum significantly modified by a forest of emission
and absorption features, possibly even P~Cygni profiles. 

Guided by our
analysis of the {\it EUVE\/} spectra of U~Gem \cite{Long96} and
OY~Car \cite{Mauche00},
we searched for identifications for the
apparent emission features in the extreme-ultraviolet 
spectrum of SS~Cyg among the
\scite{Verner96}
list  of permitted resonance lines.
This is an appropriate lamp post to look under if the lines in the 
extreme-ultraviolet
spectrum of SS~Cyg are formed predominantly by scattering in a
photoionized plasma. Identifications were established with the criteria
that the lines lie within $\Delta\lambda\approx 0.2$~\AA \ of the peaks
in the SW count spectrum, that they have large products of elemental
abundance times oscillator strength, and that they arise from a
consistent range of ionization stages. The identifications include
transitions of 
O\,VI, Ne\,V--Ne\,VIII, Mg\,V--Mg\,VII, Si\,VI--Si\,VII and Fe\,VII--Fe\,X:
intermediate charge stages of the abundant
elements. A detailed analysis of this spectrum is beyond the scope of
this paper, but we are pursuing a detailed study of the soft X-ray
spectrum of SS~Cyg with our $\lambda =40$--130~\AA \ {\it Chandra\/}
LETG spectrum 
(Mauche 2003, in preparation).

Although Fig.\,\ref{fig-euve_spec} demonstrates that the 
extreme-ultraviolet spectrum
of SS~Cyg is not well described by an absorbed blackbody, it is useful to
{\it parameterize\/} it in that way in order to obtain an estimate of
flux emitted outside of the {\it EUVE\/} bandpass. For the absorption
cross section, we used the parameterization of 
\scite{Rumph94}
for 
H\,I, He\,I and He\,II
with abundance ratios
of 1:0.1:0.01, as is typical of the diffuse interstellar medium. We
``fitted'' the observed spectrum to this model using the SW hardness
ratio $H/S$ and a fiducial count rate $H+S=0.5~\rm counts~s^{-1}$ 
to constrain $kT$, $N_{\rm H}$, and $f$ assuming $M_{\rm wd}= 1~\rm
M_{\odot}$ (hence $R_{\rm wd}=5.5\times 10^8~\rm cm$), and $d=100$ pc.
The count spectrum of one such model (with $kT=20$ eV, $N_{\rm H}
=7.6\times 10^{19}~\rm cm^{-2}$, $f=4.5\times 10^{-3}$, and $L_{\rm
bb}=2.7\times 10^{33}~\rm erg~s^{-1}$ for the observed SW count rate
$H+S=0.28~\rm counts~s^{-1}$) is shown by the smooth curve in 
Fig.\,\ref{fig-euve_spec}.

\begin{figure}
\begin{center}
\includegraphics[width=8.5cm]{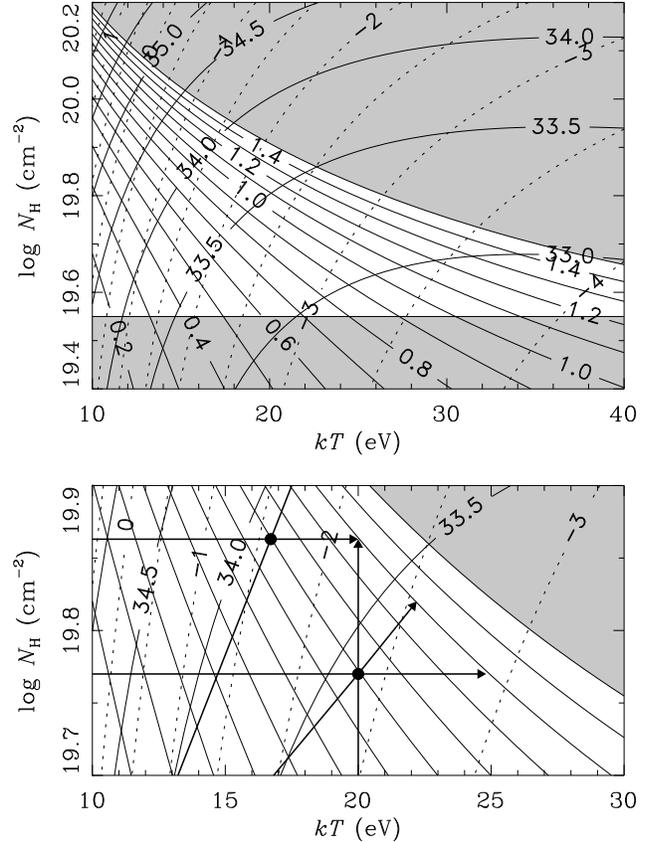}
\caption{\label{fig-euve_contour}
Constraints on the \euve\ SW spectrum of SS~Cyg.
Contours of $H/S$ ({\it solid curves\/} 0.2--1.5), $\log f$
({\it dotted curves\/} $-4$--1.5), and $\log L_{\rm bb}$ ({\it solid
curves\/} 33--36) are plotted as a function of $kT$ and $N_{\rm H}$. 
The region of
parameter space shaded gray is excluded by the conditions $H/S\le 1.5$
and $N_{\rm H}\gax 3.5\times 10^{19}~\rm cm^{-2}$. The lower panel shows
a detail of the parameter space with five example trajectories of
SS~Cyg during the course of the 1996 October outburst; the tracks
start at the filled circle and move left and/or up, then to the right
and/or down.
}
\end{center}
\end{figure}

As discussed by 
\scite{Mauche95},
it is possible to trade off $kT$ and $N_{\rm H}$
to produce acceptable fits to {\it EUVE\/} spectra of SS~Cyg, and
Figure\,\ref{fig-euve_contour} shows the parameter contours 
for $H/S$, $f$, and $L_{\rm bb}$
for $kT=10$--40 eV. A stringent upper bound on the parameter space is
set by the observed hardness ratio 
$H/S\leq$1.5, 
while a weaker lower
bound is set by the interstellar column density, which 
\scite{Mauche88}
estimated to be $N_{\rm H}\approx 3.5\times 
10^{19}~\rm cm^{-2}$ based on the curve-of-growth of ultraviolet
interstellar absorption lines. 
\scite{Mauche95}
mooted solutions with $kT=20$ eV
and 30 eV for the 1993 {\it EUVE\/} SW spectrum of SS~Cyg, 
\scite{Ponman95}
found that solutions with $kT\approx 20$--25 eV described
the {\it ROSAT\/} PSPC spectra, and we find that our {\it Chandra\/}
LETG spectrum is consistent with $kT\approx 25$ eV. Consequently, at
the peak of the outburst we favour solutions with $kT\approx 20$--25
eV, which Fig.\,\ref{fig-euve_contour} shows 
require $N_{\rm H}\approx 5.9$--$7.3\times
10^{19}~\rm cm^{-2}$ and 
$L_{\rm bb}\approx 1.7$--$4.3\times10^{33}\rm \,(d/100\,pc)^2\,erg\,s^{-1}$.

To investigate the temporal evolution of the extreme-ultraviolet 
spectrum of SS~Cyg
during its 1996 October outburst, we show in the lower panel of 
Fig.\,\ref{fig-euve_contour} five representative trajectories 
through the ($kT$, $N_{\rm H}$)
parameter space. In the first set of examples, $N_{\rm H}$ is assumed to
be constant throughout the outburst, so the trajectories follow the
horizontal tracks. In the first case, $kT$ starts out at $\approx 20$ eV
($H/S\approx 1.0$) at the beginning of the outburst, rises to $\approx
25$ eV ($H/S\approx 1.3$) at the peak of the outburst, then falls to
$\lax 17$ eV ($H/S\lax 0.8$) at the end of the outburst. In the second
case, $kT$ starts out at $\approx 17$ eV at the beginning of the outburst,
rises to $\approx 20$ eV at the peak of the outburst, and then falls to
$\lax 15$ eV at the end of the outburst. In the third example, $kT$ is
assumed to be constant at 20 eV throughout the outburst, so the trajectory
follows the vertical track. In that case, $N_{\rm H}$ starts out at
$\approx 5.9\times 10^{19}~\rm cm^{-2}$ at the beginning of the outburst,
rises to $\approx 7.3\times 10^{19}~\rm cm^{-2}$ at the peak of the
outburst, and then falls to $\lax 3.4\times 10^{19}~\rm cm^{-2}$ at
the end of the outburst. In the second set of examples, both $N_{\rm H}$
and $kT$ are assumed to increase during outburst, so the trajectories
move along the diagonal tracks in the figure. For the five examples, the
parameters at the peak of the outburst are: $kT\approx (17$, 20, 22, 25)
eV, $N_{\rm H}\approx (10$, 7.3, 6.6, $5.9)\times 10^{19}~\rm cm^{-2}$,
$f\approx (53$, 6.9, 3.0, $1.2)\times 10^{-3}$, and $L_{\rm bb}\approx
(17$, 4.3, 2.7, $1.7)\times 10^{33}\,\rm (d/100\,pc)^2\,erg\,s^{-1}$. 

Figs.\,\ref{fig-euve_hard}\,\&\,\ref{fig-euve_contour} show that
during the rise to and decline from outburst, the {\it
EUVE\/} DS and SW count rate evolution can be driven by variations in
$kT$, $N_{\rm H}$, and/or $f$ (hence $L_{\rm bb}$). At one extreme, it
is possible that $N_{\rm H}$ is fixed, and that the dramatic increase in
the count rates at the beginning of the outburst is caused mostly by an
increase in $kT$ from $\lax 10$ eV to $\approx 25$ eV. At the other
extreme, it is possible that $kT$ is fixed, and that the rapid increase
in the count rates at the beginning of the outburst is caused mostly by
an increase in $f$, hence $L_{\rm bb}$. These two extreme possibilities
imply very different bolometric corrections to convert from the observed
SW count rate to luminosity, but unfortunately we do not have the data
to determine which applies (however, see Sect.\,\ref{sect-both_spec}).

\subsection{X-ray spectra}
\label{sect-xte_spec}
\begin{figure}
\begin{center}
\includegraphics{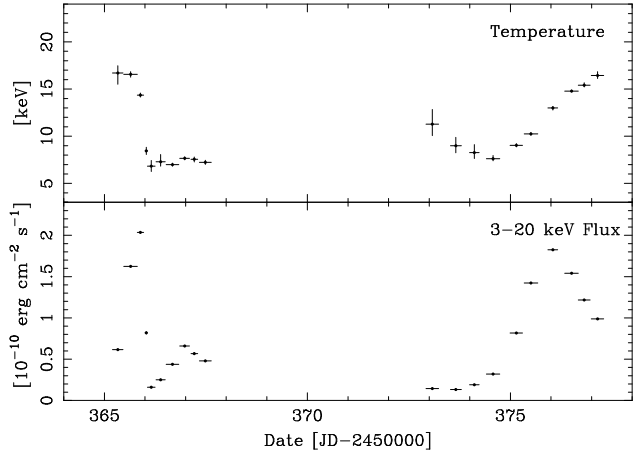}
\caption{\label{fig-xte_fit}
The results of fitting a thermal plasma model to our X-ray
spectra of SS~Cyg. 
This model consists of a {\it mekal} plasma model, 
a  6.4\,keV emission line, and an 8\,keV absorption edge. 
A full description and motivation is given by 
Wheatley (2003);
the fitted fluxes are discussed in Sect.\,\ref{sect-xte_spec}. 
All errors are 68\% confidence 
for one parameter of interest ($\Delta\chi^2=1.0$). 
}
\end{center}
\end{figure}

The \rxte\ PCA provides low-resolution X-ray spectroscopy throughout our
observations, allowing us to convert the \rxte\ count rate to X-ray
energy flux. 
The detailed analysis of the X-ray spectrum is beyond the scope of
this paper, and is presented by Wheatley (in preparation). 
We find
that the \rxte\ spectrum throughout
outburst is well represented by a single-temperature thermal-plasma
model with the addition of a 6.4\,keV line and 8\,keV absorption
edge ($\chi^2$ of 670 with 819 degrees of freedom). 
Figure\,\ref{fig-xte_fit} shows the best-fitting temperatures and fluxes. 

Our fitted fluxes provide an estimate of the accretion rate through the
boundary layer onto the white dwarf. In contrast to the \euve\ band, 
bolometric corrections are not a serious problem in hard X-rays
because the bremsstrahlung spectrum is relatively flat, the \rxte\ 
bandpass is so wide, and 
photoelectric absorption is much less strong. 

We are most interested in the accretion rates before the X-ray rise and 
immediately before the sharp X-ray suppression. The first provides an
estimate of the accretion rate in quiescence, the second provides an
estimate of the accretion rate at which the boundary layer becomes
optically-thick to its own emission. A simple inspection of 
Fig.\,\ref{fig-xte_fit} shows that the flux after the X-ray
recovery (at the end of outburst) is the same as that 
immediately before the sharp X-ray suppression. We can thus be confident
that this represents the maximum luminosity of the optically-thin
boundary layer in SS~Cyg.

The 3--20\,keV fluxes are 
$\rm 6.1\times10^{-11}\,erg\,cm^{-2}\,s^{-1}$ (from our
first spectrum)
and
$\rm 20\times10^{-11}\,erg\,cm^{-2}\,s^{-1}$ (from our
third spectrum). 
Including bolometric corrections of 1.8 
to account for flux lost outside our bandpass
we find luminosities of 
$\rm 1.3\times10^{32}\,(d/100\,pc)^2\,erg\,s^{-1}$ and
$\rm 4.4\times10^{32}\,(d/100\,pc)^2\,erg\,s^{-1}$, 
which scale to luminosities of 
$\rm 3.6\times10^{32}\,erg\,s^{-1}$ and
$\rm 12\times10^{32}\,erg\,s^{-1}$
using the HST/FGS parallax distance of 166\,pc \cite{Harrison99}.
Using the relation
$L=GM_{\rm wd}\dot{M}/2R_{\rm wd}$, and 
assuming $M_{\rm wd}= 1\,\rm M_{\odot}$ 
(hence $R_{\rm wd}=5.5\times 10^8~\rm cm$), these luminosities
correspond to accretion rates of 
$\rm 3\times10^{15}\,g\,s^{-1}$ and 
$\rm 1\times10^{16}\,g\,s^{-1}$.

Assuming the flux measured in our first spectrum
represents the true
quiescent flux, then the derived accretion rates are uncomfortably
high for disc-instability models. Predicted accretion rates in
quiescence are in the region $\rm 10^{13}\,g\,s^{-1}$
\egcite{Hameury00}, 
two and a half orders of magnitude less than that observed in SS~Cyg. 
This difference has not been satisfactorily explained, and remains one
of the main deficiencies of the disc instability model. 

In SS~Cyg the transition of the boundary layer to its
optically-thick state happens at an accretion rate only a factor three
above the quiescent rate. This is precisely the rate
predicted by \scite{Pringle79} and \scite{Patterson85}, 
though it should be pointed out that these models were developed in
response to the discovery of hard X-rays from SS~Cyg itself. 

\subsection{Combination of {\it RXTE} and {\it EUVE} spectral results}
\label{sect-both_spec}
In Sect.\,\ref{sect-lcs} we point out that the early and late \euve\
lightcurve follows the \rxte\ lightcurve closely, 
and we conclude that at those times the \euve\
count rate is dominated by photons from the soft tail of the hard
X-ray emission component. We can test this conclusion by extrapolating
our \rxte\ spectral fits to the \euve\ bandpass. 

Taking our fitted model for the third \rxte\ spectrum 
(Fig.\,\ref{fig-xte_fit}) and folding it 
through the response of the \euve\
DS instrument we find a predicted count rate of
0.2\,s$^{-1}$. Remarkably this is substantially {\em above} the measured DS
count rate at this time, 0.03\,s$^{-1}$. 
To reduce the model-predicted count rate to this level we need to
increase the absorption column density from the interstellar value of 
$N_{\rm H}=3\times10^{19}\rm \,cm^{-2}$ to $13\times10^{19}\rm
\,cm^{-2}$. This increase is acceptable since the higher absorption
still results in negligible absorption in the \rxte\ bandpass. 
Assuming the absorption does not vary during outburst, this high
$N_{\rm H}$ points to the higher luminosity fits to the \euve\
spectrum in Sect.\,\ref{sect-euve_spec}: around
$\rm 10^{35}\,erg\,s^{-1}$ at maximum, corresponding to accretion rates around 
$\rm 10^{18}\,g\,s^{-1}$ (assuming $M_{\rm wd}=1\,M_\odot$ and d=166\,pc). 

A second constraint on the \euve\ spectrum from our \rxte\ fitting is
the boundary layer flux at the time of the transition from dominant 
X-ray to extreme-ultraviolet emission. Our \rxte\ fitting shows that
the X-ray emission reaches a maximum flux of 
$\rm 2\times10^{-10}\,erg\,cm^{-2}\,s^{-1}$. Assuming that the accretion
rate continues to rise during this transition, the
extreme-ultraviolet spectrum must be capable of radiating at the same
rate, but with a measured count rate of just 0.01\,s$^{-1}$ 
(see Fig.\,\ref{fig-all}). At $N_{\rm H}=3\times10^{19}\rm \,cm^{-2}$
this implies a blackbody temperature of just 6\,eV.  Even with the
higher absorption implied by the extrapolation of the \rxte\ spectrum
to the \euve\ band ($N_{\rm H}=1.3\times10^{20}\rm \,cm^{-2}$) 
the blackbody temperature can rise only to 9\,eV.
Thus looking again at the possible evolutionary tracks of the 
extreme-ultraviolet
section outlined in Sect.\,\ref{sect-euve_spec} and 
Fig.\,\ref{fig-euve_contour}, we find that the blackbody temperature must
begin low and increase during the extreme-ultraviolet rise. 
We can thus rule out spectral evolution in which only the absorption varies.

\section{Discussion} 

\subsection{The X-ray delay}
\label{sect-dis_delay}
For the first time we have resolved the very beginning of the
high-energy rise in a dwarf nova outburst. The rise begins in hard
X-rays 0.9--1.4\,d after the beginning of the rise in the optical, and
0.5\,d after the optical brightness crosses $\rm m_{vis}$=11 (which is
better defined). The extreme-ultraviolet emission does not begin to
rise for another 0.6\,d. 

We interpret the beginning of the X-ray rise as the moment at which outburst
material first reaches the boundary layer. It thus provides a precise
measurement of the time the outburst heating wave reaches the
boundary layer. Unfortunately the optical observations are not
sufficiently evenly spaced to allow us to determine the precise timing
of the beginning of the progression of the heating wave, but the
propagation time must be in the range 0.9--1.4\,d. 

The first measurements of a delay between the rise at long and short
wavelengths was made with the {\it International Ultraviolet Explorer}
\egcite{Hassall83}, and it has thus been known as the ``ultraviolet
delay''.
However, we believe that the {\em X-ray delay} is a better measure of
the propagation time of disc heating wave. This is because X-rays are
emitted from a precisely known location at the inner edge of the
accretion disc, as determined by eclipse studies \cite{Mukai97,Wheatley03-OY}.

Target-of-opportunity observations with \euve\ have yielded precise
rise times for three dwarf novae \cite{Mauche01b} and several authors
have used these as a measure of the heating wave
propagation times \egcite{Smak98,Hameury99,Stehle99,Cannizzo01,Mauche01b}.
Our result indicates that the heating wave must propagate even faster
than inferred by these authors.

\subsection{Boundary layer emission}
\label{sect-dis_bl}
At the beginning of the high-energy outburst the X-ray emission rises sharply, 
but with no change in the X-ray spectrum. We take this as evidence that the 
quiescent and early-outburst X-ray emission arise from the same source, 
presumably the boundary layer. Furthermore, the precise co-incidence of the 
timing of the EUV rise and the X-ray suppression shows that the EUV component 
must also arise in the boundary layer. 

Our measured X-ray luminosities (Sect.\,\ref{sect-xte_spec}) 
translate to accretion rates of 
$\rm 3\times10^{15}\,g\,s^{-1}$ 
in quiescence and
$\rm 1\times10^{16}\,g\,s^{-1}$
immediately before the boundary layer becomes
optically-thick. This quiescent rate is two and a half orders of
magnitude higher than that predicted by the standard disc instability model 
\egcite{Hameury00}. Somehow
the models need to allow the quiescent disc to transfer more
material. 

One way to increase the accretion rate of dwarf novae in quiescence 
is to truncate the accretion disc. 
Several mechanisms have been proposed including 
the weak magnetic field of the white dwarf \cite{Livio92}, some kind of 
heating mechanism leading to disc evaporation \cite{Meyer94}, 
and irradiation by the hot white dwarf \cite{King97}. 
\scite{Done97} claim to see the effects of this truncation in the 
X-ray spectrum of SS~Cyg (but see also Wheatley \& Mauche, in prep.). 

Rather than truly truncating the quiescent accretion disc, the scheme of 
\scite{King97} forces in the inner disc to remain in the outburst state 
permanently. Since material is transferred efficiently in this state, the 
resulting accretion rate is the same as that of a truncated disc. 
The calculations of \scite{Truss00} suggest that the inner accretion disc can 
remain in the outburst state through quiescence, even without the need for 
irradiation. This would naturally explain the high X-ray luminosity of SS~Cyg 
in quiescence.

\subsection{Origin of the residual X-ray emission during outburst}
\label{sect-dis_xob}
Although we are confident that 
the hard X-ray emission during quiescence and the rise at the beginning 
of outburst are due to the boundary layer, 
we are less sure of the origin of the X-ray emission
during outburst. This emission has a much lower temperature, which is
puzzling if one believes the temperature is set by the depth of the potential
well and not the accretion rate. It is also a much smaller fraction of
the bolometric luminosity. 

\scite{Patterson85} suggest that the
outburst emission is due to a density gradient in the outburst boundary layer,
such that the upper levels of this region are optically thin. 
Their Figure\,8 shows a schematic in which an optically-thin hard X-ray
emitting region sits above the optically-thick
extreme-ultraviolet-emitting boundary layer. 
Their interpretation is
attractive because it does not require a second source of the X-rays. 

On the other hand, eclipse observations of OY~Car do provide evidence for a 
second source of X-ray emission during outburst. 
X-ray observations \cite{Naylor88,Pratt99b} and extreme-ultraviolet
\cite{Mauche00} observations of OY Car in superoutburst 
show that the X-ray and extreme-ultraviolet emitting
regions must be much larger during outburst than in quiescence. 
\scite{Mauche00} argue that the {\it EUVE\/} extreme-ultraviolet 
spectrum of OY~Car in
superoutburst is due to scattering of optically-thick
boundary layer emission in an extended photo-ionised accretion-disc wind. 
The direct boundary layer emission must by obscured from 
view, probably by the disc itself. 

In contrast, scattering is unlikely to account for the extended source of 
X-rays in OY Car in superoutburst because of the relative absence of strong
resonance lines in the X-ray band. Any scattered component must therefore be 
much weaker even than the direct 
flux visible in low-inclination systems such as SS~Cyg.  
{\it ROSAT\/} observations, however,  show that OY~Car is actually
brighter in outburst than quiescence. \scite{Pratt99a} find a {\it ROSAT\/}
HRI count rate of 0.016\,s$^{-1}$ in quiescence and \scite{Pratt99b}
find an outburst count rate of 0.028\,s$^{-1}$ (note that the
quiescent count rates in their Table\,1 refer to the PSPC detector). 
We conclude that
the relative brightness of the outburst X-ray emission of OY~Car points to a 
second source of X-ray emission that must be inherently larger than the
boundary layer.  

A candidate is a magnetically-heated accretion-disc corona. 
This is plausible because the accretion disc is believed to be more 
magnetically active in outburst than in quiescence, and indeed it is a 
magneto-rotational instability that is believed to drive the 
enhanced effective viscosity during outburst \cite{Balbus91}.

\section{Conclusions}
In this paper we present simultaneous optical, extreme-ultraviolet
and X-ray observations of SS~Cyg throughout outburst. For the first
time we resolve the beginning of the high-energy outburst, which 
starts
in the hard X-ray band
0.9--1.4\,d after the beginning of the optical 
rise (Sect.\,\ref{sect-lcs}). 
The X-ray rise continues for 0.6\,d with the
luminosity increasing from $\rm 3.6\times10^{32}\,erg\,s^{-1}$ 
to $\rm 1.2\times10^{33}\,erg\,s^{-1}$. The implied accretion rate in 
quiescence is two and a half orders of magnitude higher than predicted
by the standard disc instability model, 
and this remains one of the main deficiencies of that model
(Sects.\,\ref{sect-xte_spec}\,\&\,\ref{sect-dis_bl}).
This discrepancy may be resolved if one allows the inner accretion disc 
to remain in the outburst state throughout quiescence \egcite{Truss00}.
After 0.6\,d the X-ray flux drops to near zero in less than three
hours, and is replaced immediately with extreme-ultraviolet emission. 
This is the first time this transition has been resolved, and the
coincidence of the X-ray fall and extreme-ultraviolet rise leaves no
doubt that both components originate in the boundary layer between accretion
disc and white dwarf. 
The boundary layer may also account for the residual hard X-ray emission in 
quiescence, but we argue that some form of magnetic heating of a 
disc corona is more likely (Sect.\,\ref{sect-dis_xob}). 

The X-ray rise defines the moment at which outburst material 
first reaches the boundary layer, and 
so authors who have used the extreme-ultraviolet rise
to infer the propagation time of the outburst heating wave may have
over-estimated the propagation time by 0.6\,d in SS~Cyg 
(Sect.\,\ref{sect-dis_delay}).

The extreme-ultraviolet band dominates during most of the outburst,
peaking at a luminosity around $\rm 5\times10^{34}\,erg\,s^{-1}$
(Sects.\,\ref{sect-euve_spec}\,\&\,\ref{sect-both_spec}). 
Dwarf nova oscillations are seen with
high amplitude throughout the extreme-ultraviolet outburst, but are
not present in our X-ray lightcurves. This
suggests that the source of the oscillations operates only in the
optically-thick boundary layer, though it is also possible that the
X-ray oscillations are smeared out by a long X-ray cooling timescale
(Sect.\,\ref{sect-dnos}). 

At the end of outburst the boundary layer emission switches back to the
X-ray band, this time more slowly, but at the same
luminosity/accretion rate as before (Fig.\,\ref{fig-xte_fit}). 
The instrumental count rates clearly depend strongly on the changing
spectrum during these transitions, and the X-ray hardness ratio
appears to be the best indicator of the accretion 
rate at these times (Sect.\,\ref{sect-hard_xte}).  
The transitions between X-ray and extreme-ultraviolet emission are
accompanied by intense variability, that (sometimes at least) takes the
form of quasi-periodic oscillations with periods around 200\,s 
(Sect.\,\ref{sect-var}). 
These variations are not associated with hardness variations and 
by inferring that the X-ray cooling timescale must be shorter than 
these variations we were
able to limit the density of the X-ray emitting plasma to be greater than
$5\times10^{13}\,\rm cm^{-3}$ (Sect.\,\ref{sect-cool}).

\section*{Acknowledgments}
We thank 146 AAVSO observers worldwide who
closely monitored the October 1996 outburst of SS Cyg and contributed
784 optical observations to the AAVSO International Database. 
Their contribution provided the vital trigger for the satellite
observations and allowed us to correlate the X-ray and the EUV data.
The rapid-response {\it RXTE\/} observations were made possible 
by \rxte\ Project Scientist J.\ Swank; \rxte\ Chief Mission 
Planner E.\ Smith; and the staff of the \rxte\ Science Operations
Center at the Goddard Space Flight Center.
{\it EUVE\/} observations were made possible by {\it EUVE\/} Deputy
Project Scientist R.\ Oliversen; {\it EUVE\/} Science Planner B.\
Roberts; the staff of the {\it EUVE\/} Science Operations Center at CEA,
and the Flight Operations Team at Goddard Space Flight Center.
We thank the referee Guillaume Dubus for his thorough reading of this 
paper and for many useful and constructive comments. 
PJW acknowledges funding from PPARC through rolling grants held at the 
University of Leicester. 
CWM's contribution to this work was performed under the auspices
of the U.S.\ Department of Energy by University of California Lawrence
Livermore National Laboratory under contract No.\ W-7405-Eng-48.
JAM gratefully acknowledges NASA grant
NAG5-7027 for AAVSO participation in this project.

\bibliographystyle{mnras3}
\bibliography{mn_abbrev2,refs}

\end{document}